\documentclass[prl,twocolumn]{revtex4}

\begin{document}
\title{Comment on ''Theory of tailoring sonic devices: Diffraction dominates over
refraction''}

\date{\today}

\author{A. H\aa kansson,$^{1}$ J. S\'{a}nchez-Dehesa,$^{1,*}$ F.Cervera,$^{2}$F. Meseguer$^{1,2}$,
 L.Sanchis$^{3}$, and J. Llinares$^{2}$}

\affiliation{$^{1}$Centro de Tecnolog\'{\i}a Nanofot\'{o}nica and Dpto.Ingenier\'{\i}a
Electr\'{o}nica,\\
Universidad Polit\'{e}cnica de Valencia, C/ Camino de Vera s/n, E-46022 Valencia, Spain.}

\affiliation{$^{2}$Centro Tecnol\'{o}gico de Ondas, Unidad Asociada de Investigaci\'{o}n
(CSIC-UPV), Edificio de Institutos II, \\
Universidad Polit\'{e}cnica de Valencia, C/ Camino de Vera s/n, E-46022 Valencia, Spain.}

\affiliation{$^{3}$Departamento de F\'{\i}sica Te\'orica de la Materia Condensada, Facultad de
Ciencias (C-V),\\ Universidad Aut\'{o}noma de Madrid, E-28049 Madrid, Spain.}

\begin{abstract}

Recently N. Garc\'{\i}a {\it et al}. (Phys. Rev. E {\bf 67}, 046606 (2003))
theoretically studied several acoustic devices with dimensions on de order
of several wavelenghts. The authors discussed experimental results
previously reported by several of us (F. Cervera {\it et al}., Phys. Rev.
Lett. {\bf 88}, 023902 (2002)). They concluded that it is diffraction rather than 
refraction that is the dominating mechanism explaining the focusing effects observed
in those experiments. In this Comment we reexamined their calculations and discussed
why some of their interpretations of our results are misleading.

\end{abstract}


\maketitle

The recent paper by Garcia {\it et al}. \cite{NGarcia} addressed an issue of
interest in the field of acoustic crystals (ACs). It concerns the role that
diffraction plays vs. refraction in determining the effects observed in
acoustic devices with dimensions of the order of several wavelengths. In our
opinion, this issue is related with the problem of homogenization of
clusters consisting of periodic arrangements of sonic scatterers in air. In
other words, if the AC-based device is large enough so that its properties
can be explained in terms of an effective medium theory (where a refractive
index can be defined), one would say that refraction dominates over
diffraction. The existence of a critical size above one can consider that
refraction dominates over diffraction is an issue that was not taken into
account in the paper by Garcia {\it et al}. \cite{NGarcia}.

In regards with the acoustic devices presented in Ref.\cite{NGarcia}, we
agree to the general conclusion obtained by the authors from their
theoretical simulations; i.e., focusing phenomena and image formation are
dominated by diffraction rather than refraction due to the small dimension
of the acoustic devices studied. Nevertheless, the authors in Ref. \cite
{NGarcia} criticize the results recently reported by several of us for much
larger structures, for which we claimed that refraction is a dominant
mechanism.\cite{FCervera}. This Comment is to clarify on some misconceptions
and criticisms made by the authors of Ref. \cite{NGarcia}. We also have
reexamined their predictions and new experiments will be presented that
confirm our own simulations based on multiple scattering theory (MST).

In order to reproduce experimental findings, Garcia {\it et al}. \cite
{NGarcia} used acoustical devices like those reported in \cite{FCervera} but
with much smaller sizes. As a first case, they employed a FDTD method to
simulate the sound scattering by a biconvex cylindrical lens made of only 32
aluminum rods, which they claim ''{\it is similar to that of experiment in
Ref. [6]}'' (Ref.\cite{NGarcia} in this Comment). \ In this regard, we have
to comment that the actual size of the crystal lens employed in our
experiment is about 6 times bigger, which has a crucial difference when an
analysis of refraction vs. diffraction is made. Figure 1(a) shows the
comparison between both structures. As a second case, Ref. \cite{NGarcia}
presented the simulation of the sound scattering by a slab consisting of
only 28 rods to support that focusing effects is dominated by diffraction.
At this point, we have to remark that the actual slab employed in our
experiments consists of 400 aluminum rods (see Fig. 4 in Ref. \cite{FCervera}%
). A comparison between both slabs is shown in Fig. 1(b). Obviously, these
big differences between the structures theoretically modeled and the ones
experimentally employed, made completely misleading the comparison between
theory and measurements. Therefore, the smaller size of the structures does
not support the argumentation made by Garcia {\it et al}. In our opinion,
the pressure maps shown in Fig. 4 of Ref. \cite{FCervera} clearly
demonstrated our conclusion that our lens is dominated by refraction rather
than diffraction. Diffraction effects, although present at the edge zones,
are completely negligeable. On this concern, a theoretical discussion about
acoustic lens have been recently reported by Gupta and Ye \cite{BCGupta},
who used MST to perfectly reproduce our measurements. A further support of
the fact that refraction and not diffraction is the dominating mechanism in
clusters of comparable size has been recently presented by some of us in
Ref. \cite{LSanchis}, which demonstrated the homogenization of crystal slabs
with dimensions similar to the ones used in Ref. \cite{FCervera}.

If the acoustic device has a number of scatterers as low as those modeled by
Garc\'{i}a {\it et al}., we completely agree that diffraction is the
dominant mechanism. To support this conclusion, we made our own theoretical
simulations by means of MST as well as measurements on the same structures
studied in Ref. \cite{NGarcia}. Figures 2(a) and 2(b) show that our
theoretical simulations are in agreement with the measurements. At this
points, let us remark that our simulations slightly differs with the ones
presented in Figs. 2(a) and 2(b) of Ref. \cite{NGarcia}. One can observe
that the focal point is located at the same distance in the two structures,
which contradict the commnent made by Garc\'{i}a et al.. The differences are
probably due to the intrinsic limitations of the FDTD method, which does not
treat exactly the scattering by a cylindrical rod as the MST does.

To conclude, an important issue is still unsolved: it concerns with the
problem of homogenization of acoustic crystals having small dimensions in
order to determine the minimum size of cluster at which its properties can
be described by effective values of its acoustical parameters.

Work partially supported by CICyT of Spain.

\newpage \narrowtext
\begin{figure}
\caption{(a) The circles (black and white) define the total set of aluminum
cylinders reported as an acoustic lens in Ref. \protect\cite{FCervera}. The
partial set defined by the black circles corresponds to the structure
employed in the simulation of an acoustic lens in Ref. \protect\cite{NGarcia}%
. (b) The circles (black and white) circles define the set of aluminum
cylinders reported as an acoustic Fabry-Perot interferometer in Ref. 
\protect\cite{FCervera}. The partial set defined by the black circles
corresponds to the structure employed in the simulations reported in Ref. 
\protect\cite{NGarcia}. The separation between ticks in both figures
corresponds to one wavelenght.}
\end{figure}

\begin{figure}
\caption{ (a) (top panel) Calculated pressure pattern (in dB) of an incident
sound plane wave (1700 Hz wavelenght) scattered by a lenslike periodic
arrangements of rigid rods (white circles) with hexagonal symmetry. (a)
(bottom panel) Measured pressure pattern of the corresponding structure made
of aluminun cylinders. (b) (top panel) Calculated pressure pattern (in dB)
of an incident sound plane wave (1700 Hz wavelenght) scattered by a
rectangular slab of rigid rods (white circles). (a) (bottom panel) Measured
pressure pattern of the corresponding structure made of aluminum rods.
Details of calculation's method and measurements can be found in Ref.
\protect\cite{LSanchis}.}
\end{figure}

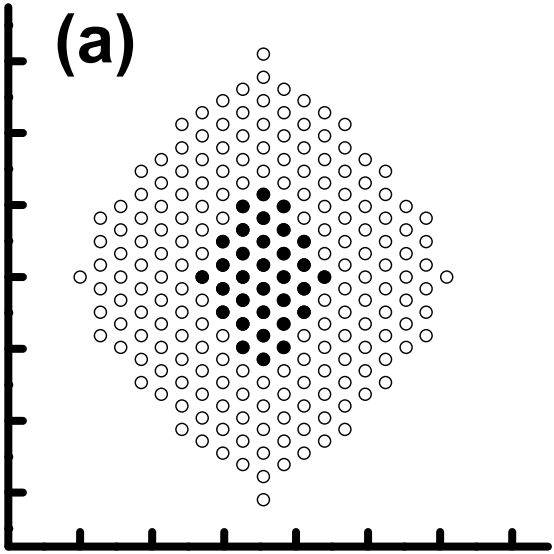
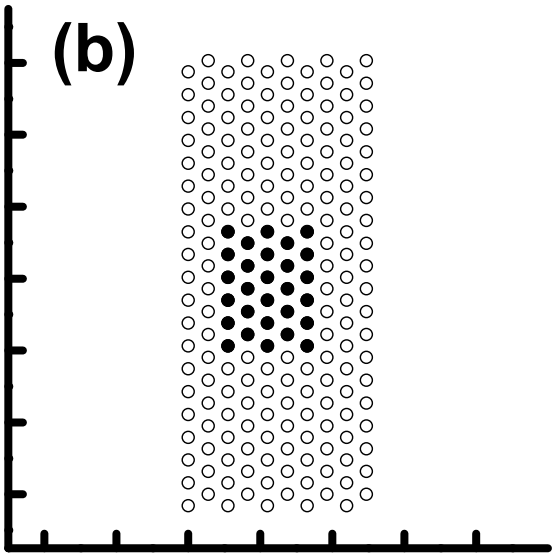
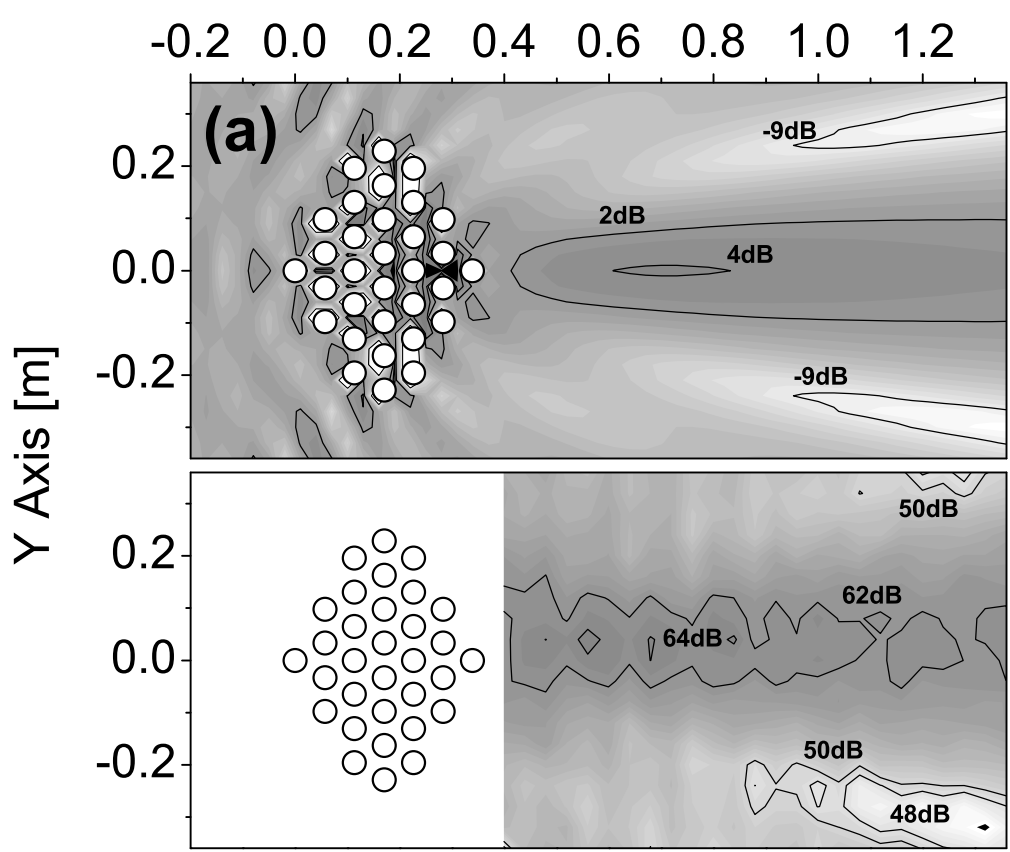
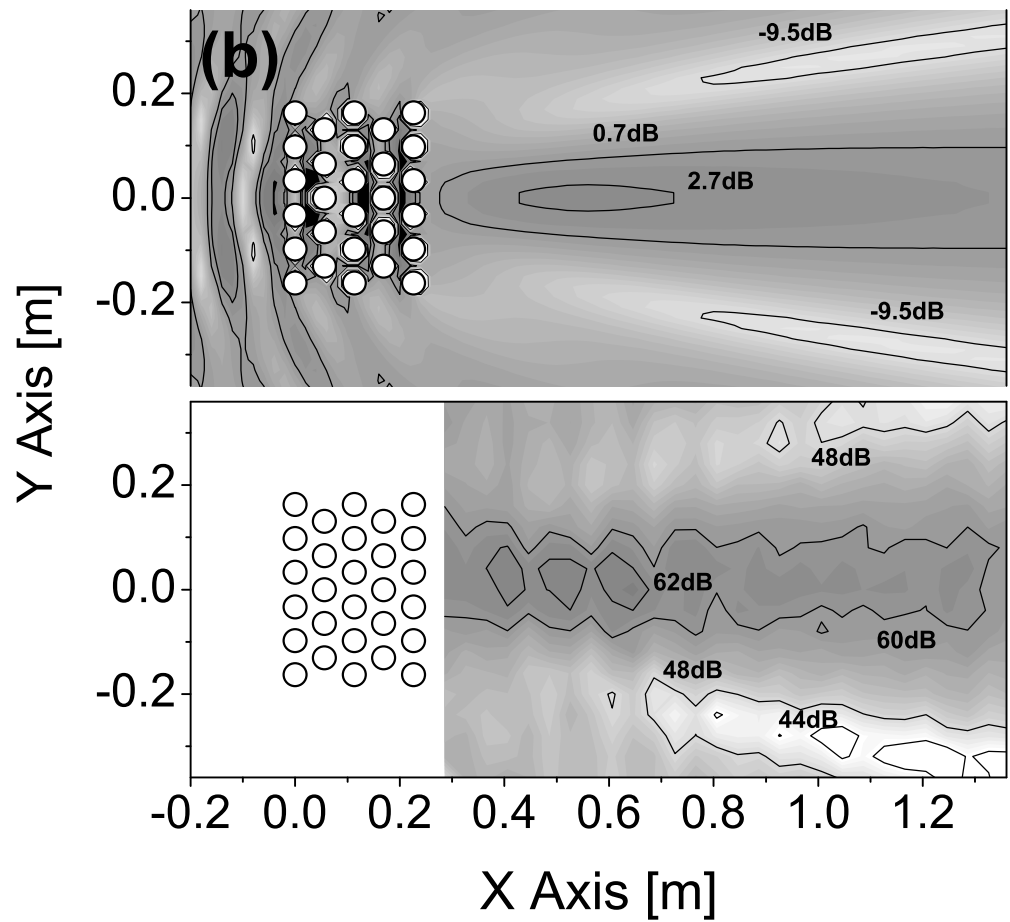
\end{document}